\documentclass{PoS}
\usepackage{amssymb}
\usepackage{comment}

\title{The Far-Infrared Radio Correlation at High-z: Prospects for the SKA}

\ShortTitle{The Far-Infrared Radio Correlation at High-z: Prospects for the SKA}

\author{\speaker{Eric Murphy}\\
       Caltech\\
       E-mail: \email{emurphy@ipac.caltech.edu}}

\abstract{I present a predictive analysis for the behavior of the far-infrared (FIR)--radio correlation as a function of redshift in light of the deep radio continuum surveys which may become possible using the square kilometer array (SKA).  
To keep a fixed ratio between the FIR and predominantly non-thermal radio continuum emission of a normal star-forming galaxy, whose cosmic-ray (CR) electrons typically lose most of their energy to synchrotron radiation and Inverse Compton (IC) scattering, requires a nearly constant ratio between galaxy magnetic field and radiation field energy densities.  
While the additional term of IC losses off of the cosmic microwave background (CMB) is negligible in the local Universe, the rapid increase in the strength of the CMB energy density (i.e. $\sim(1+z)^{4})$ suggests that evolution in the FIR-radio correlation should occur with infrared (IR;~$8-1000~\mu$m)/radio ratios increasing with redshift.  
This signature should be especially apparent once beyond $z\sim3$ where the magnetic field of a normal star-forming galaxy must be  $\sim$50~$\mu$G to save the FIR-radio correlation.  
At present, observations do not show such a trend with redshift; 
$z\sim6$ radio-quiet quasars appear to lie on the local FIR-radio correlation while a sample of $z\sim4.4$ and $z\sim2.2$ submillimeter galaxies (SMGs) exhibit ratios that are a factor of $\sim$2.5 {\it below} the canonical value.  
I also derive a 5$\sigma$ point-source sensitivity goal of $\approx$20~nJy (i.e. $\sigma_{\rm RMS} \sim 4$~nJy) requiring that the SKA specified be $A_{\rm eff}/T_{\rm sys}\approx  15000$~m$^{2}$~K$^{-1}$; 
achieving this sensitivity should enable the detection of galaxies forming stars at a rate of $\gtrsim25~M_{\odot}~{\rm yr}^{-1}$, such as typical luminous infrared galaxies (i.e. $L_{\rm IR} \gtrsim 10^{11}~L_{\odot}$), at all redshifts if present.   
By taking advantage of the fact that the non-thermal component of a galaxy's radio continuum emission will be quickly suppressed by IC losses off of the CMB, leaving only the thermal (free-free) component, I argue that deep radio continuum surveys at frequencies $\gtrsim$10~GHz may prove to be the best probe for characterizing the high-$z$ star formation history of the Universe unbiased by dust.  }

\FullConference{Panoramic Radio Astronomy: Wide-field 1-2 GHz research on galaxy evolution\\
		 June 2-5 2009\\
		 Groningen, the Netherlands}

\begin{document}

\section{Introduction}
Radio continuum emission from galaxies arises due to a combination of thermal and non-thermal processes primarily associated with the birth and death of young massive stars, respectively.  
The thermal (free-free) radiation of a star-forming galaxy is emitted from H{\sc ii} regions and is directly proportional to the photoionization rate of young massive stars.  
Since emission at GHz frequencies is optically thin, the thermal radio continuum emission from galaxies is a very good diagnostic of a galaxy's massive star formation rate.  
Massive ($\gtrsim 8~M_{\odot}$) stars which dominate the Lyman continuum luminosity also end their lives as supernovae (SNe) whose remnants (SNRs) are responsible for the acceleration of cosmic-ray (CR) electrons into a galaxy's general magnetic field resulting in diffuse synchrotron emission.  
These same massive stars are often the primary sources of dust heating in the interstellar medium (ISM) as their starlight is absorbed and reradiated at far-infrared (FIR) wavelengths by interstellar grains.  
 This common origin between the FIR dust emission and thermal $+$ non-thermal radio continuum emission from galaxies is thought to be the dominant physical processes driving the FIR-radio correlation on global (e.g. \cite{gxh85,yrc01}, and references therein) 
 and local (e.g. \cite{ejm06,ejm08}, and reference therein) scales.
At present,  all indications suggest that the FIR-radio correlation holds out to moderate redshifts (e.g. \cite{df06,ejm09a,ms09}, and references therein).  


The detection of large populations of dusty star-forming galaxies at high redshift by ISO and {\it Spitzer} has underscored the need for reliable star formation rate diagnostics unaffected by dust.
While radio emission may provide an excellent advantage over other wavelengths, 
 detecting large populations of high redshift star-forming galaxies at radio wavelengths remains extremely difficult.  
While detectable with current FIR capabilities, even with a fully operational EVLA, IR-bright star-forming galaxies  (e.g. M~82; $L_{\rm IR} \approx 4\times10^{10}~L_{\odot}$) and moderate LIRGs  (i.e. $L_{\rm IR} \approx 3\times10^{11}~L_{\odot}$) will not be detectable beyond redshifts of $z\sim 1$ and $z\sim 2$, respectively.  
A next-generation radio facility such as the Square Kilometer Array (SKA) should easily remedy this disparity between the depth of FIR and radio continuum surveys.  

Recently, \cite{ejmc09} described physically motivated predictions for the evolution of the FIR-radio correlation as a function of redshift arising from variations in the CR electron cooling time-scales as Inverse Compton (IC) scattering off of the Cosmic Microwave Background (CMB) becomes increasingly important.   
Since the non-thermal component of a galaxy's radio continuum is increasingly suppressed with redshift, radio continuum measurements at moderately high frequency ($\sim$10~GHz) become one of the cleanest ways to quantify the star formation activity of galaxies at high redshifts unbiased by dust.  
In this proceedings article I summarize some of these findings with an emphasis placed on what this might mean for deep radio continuum surveys using the SKA.

 \section{Suppression of Non-Thermal Emission by the CMB \label{sec-ntsup}}
The FIR-radio correlation relies on a fixed ratio between synchrotron and the total energy losses of CR electrons.    
At $z =0$, the radiation field energy density of the CMB is $U_{\rm CMB} \sim 4.2\times10^{-13}~{\rm erg~cm^{-3}}$, significantly smaller than the radiation field energy density of the Milky Way (i.e. $U_{\rm MW} \sim 10^{-12} ~{\rm erg~cm^{-3}}$).  
Thus, CR electron energy losses from IC scattering off the CMB are negligible at low redshifts.  
However, $U_{\rm CMB} \propto (1+z)^{4}$ making such losses increasingly important with redshift.  
For instance, by $z\sim3$, $U_{\rm CMB} \sim 1.1\times 10^{-10}~{\rm erg~cm^{-3}}$; equating this to the magnetic field energy density, $U_{B}=B^{2}/(8\pi)$, results in a corresponding magnetic field strength of $\sim$50~$\mu$G, nearly an order of magnitude larger than the ambient field strength in the Solar Neighborhood.  
Consequently, the non-thermal component of a galaxy's radio continuum emission will be increasingly suppressed with increasing redshift, eventually resulting in only the thermal component being detectable.  

Assuming an intrinsic FIR-radio correlation for galaxies with $q_{\rm IR} = \log(F_{\rm IR}/3.75\times10^{12}) - \log (S_{\rm 1.4~GHz})\approx 2.64$ \cite{efb03}, the expected observed-frame 1.4~GHz flux density for star-forming galaxies having a range of IR ($8-1000~\mu$m) luminosities as a function of redshift are estimated (Figure \ref{fig-1}a).     
It is assumed that the radio continuum emission is comprised of two components, thermal (free-free) and non-thermal (synchrotron) emission, both of which can be expressed as power-laws such that $S_{\nu} \propto \nu^{-\alpha}$, where $\alpha_{\rm T} \sim 0.1$ for the thermal (free-free) component and $\alpha_{\rm NT} \sim 0.8$ for the non-thermal component.  
A thermal fraction of $\sim$10\% at 1.4~GHz is also assumed.  
Increased energy losses to CR electrons from IC scattering as $U_{\rm CMB}$ increases with redshift are explicitly included along with a $(1+z)^{-1}$ correction factor to account for bandwidth compression.  
Figure \ref{fig-1}a illustrates two cases per IR luminosity in which the galaxy has a 10 and 100~$\mu$G field, indicated by the solid and dotted lines, respectively.     

For the case of a moderately bright LIRG (highlighted in Figure \ref{fig-1}a), it is clearly shown that its detection in the radio continuum may significantly rely on the strength of its magnetic field between $1 \lesssim z \lesssim 6$ due to the suppression of its synchrotron emission by increased IC losses off of the CMB.  
At higher redshifts it appears that the strength of the magnetic field plays less of a role since IC losses off of the CMB begin to completely suppress a galaxy's non-thermal emission, making only its thermal radio emission detectable.  
It is also shown that a moderately bright LIRG at $z\sim10$ should have a 1.4~GHz flux density of $\sim$40~nJy almost entirely arising from free-free emission.  
Using the star formation rate conversion of \cite{rk98}, this implies that all galaxies forming stars at a rate of  $\gtrsim50~M_{\odot}~{\rm yr}^{-1}$ should be detectable out to $z\sim10$ if such a population of galaxies exists at these early epochs and this sensitivity is achievable.  


\subsection{Expected Evolution in the FIR-Radio Correlation with Redshift from CMB Effects}
The suppression of a galaxy's non-thermal emission due to the increasing importance of IC scattering off of the CMB with redshift should be directly reflected in changes to a galaxy's IR/radio ratio.  
Assuming that the intrinsic IR/radio ratio of star-forming galaxies is not unique to $z\sim0$ systems, the FIR-radio correlation should evolve in a very predictable way as a function of redshift arising from the increase in the energy density of the CMB.  
This is illustrated in Figure \ref{fig-1}b along with observational results from the literature.   
When the non-thermal emission is completely suppressed by IC scattering off of the CMB, observed IR/radio ratios are expected to simply approach the ratio between the FIR and thermal radio continuum (i.e. $q_{\rm IR}$ ratios which are a factor of $\sim$10 higher than the nominal value).    

Different CR electron cooling processes may become increasingly more important than synchrotron radiation in galaxies hosting strong starbursts.  
For such cases, the non-thermal component may diminish even more quickly than what is illustrated here.  
However, it is also possible that other physical processes could work in the opposite direction to increase the synchrotron emissivity, such as an increase to the acceleration efficiency of CR electrons within SNRs, as well as additional synchrotron emission arising from an increase to secondary positrons and electrons.  
Such  possibilities are only speculative since local starbursts do not seem to exhibit such variations in their IR/radio ratios.

\subsection{Deep Continuum Surveys using the SKA}  
\begin{figure*}
\center{
\includegraphics[width=.48\textwidth]{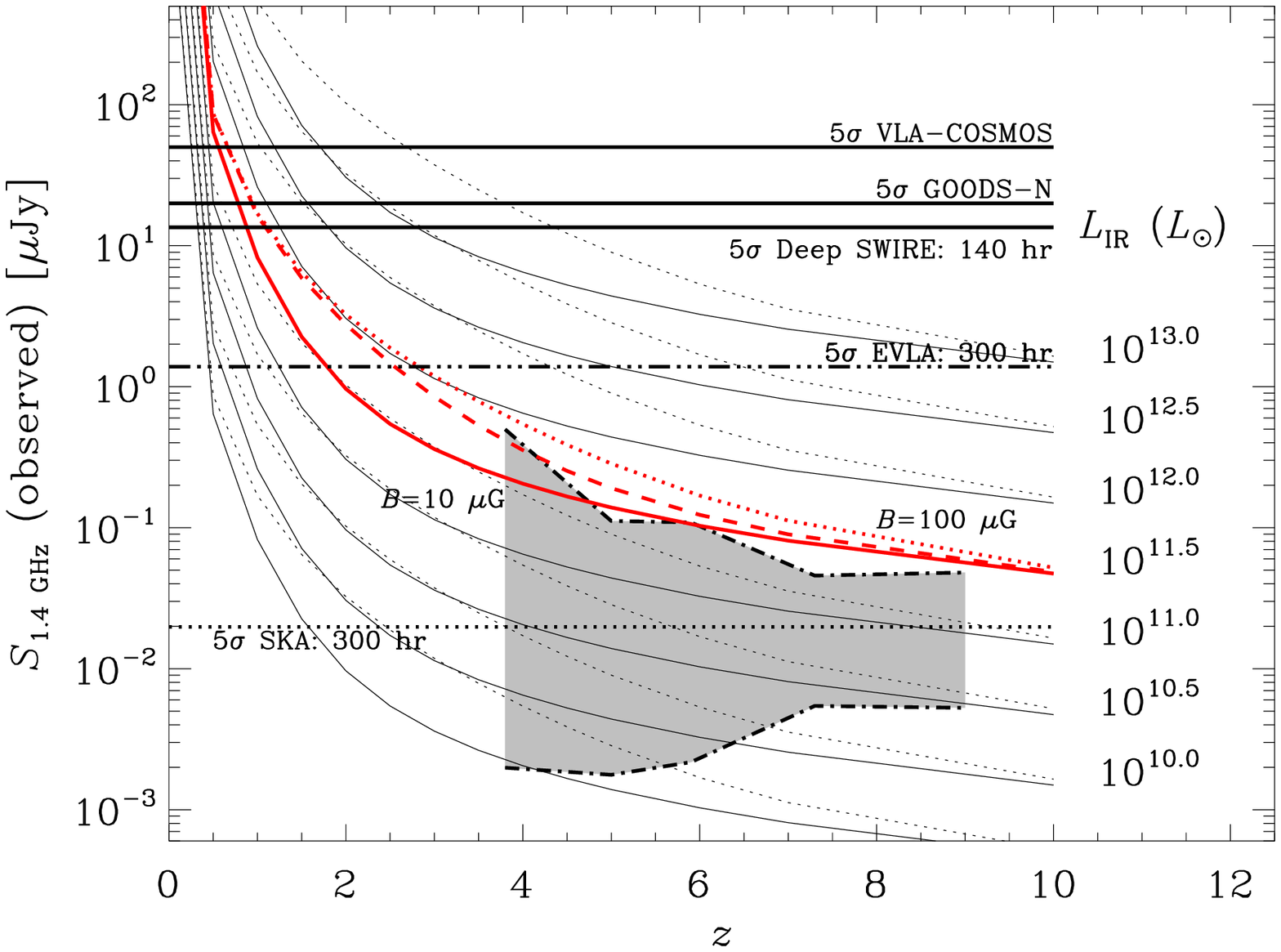}
\includegraphics[width=.47\textwidth]{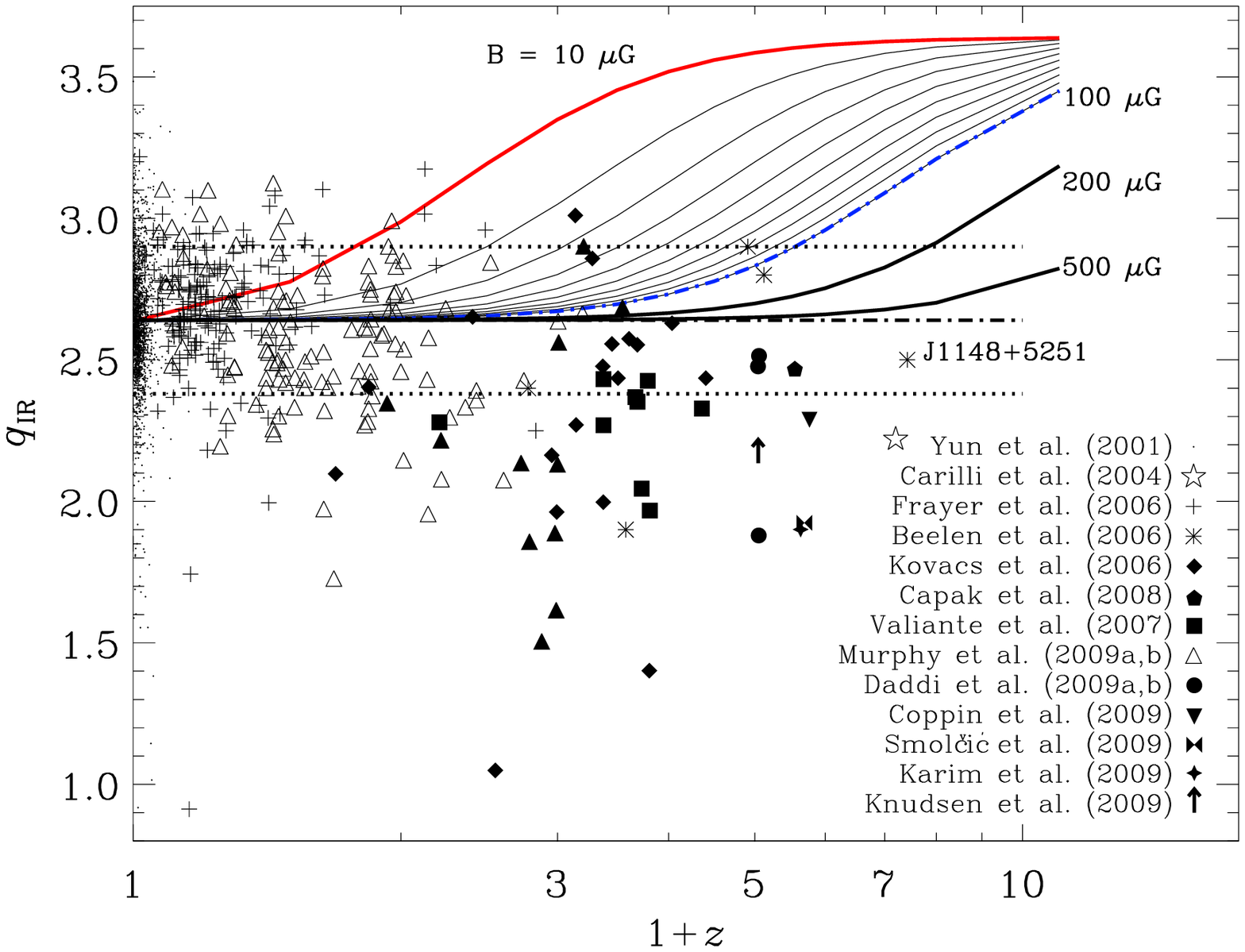}}
\caption{\scriptsize
{\it Left~(a)-} The expected observed-frame 1.4~GHz flux density for galaxies of varying IR luminosities assuming the FIR-radio correlation (i.e. $q_{\rm IR}=2.64$) as a function of redshift for nominal galaxy parameters.    
Additional energy losses to CR electrons arising from IC scattering off of the CMB, whose radiation energy density scales as $\sim(1+z)^{4}$, are included.   
Flux densities, corrected for bandwidth compression, are plotted assuming internal magnetic field strengths of 10~$\mu$G ({\it solid}-line) and 100~$\mu$G ({\it dotted}-line); 
at high~$z$ the non-thermal emission from galaxies is highly suppressed even if $B\sim100~\mu$G, leaving only the thermal component detectable.   
The case of a moderate LIRG ($L_{\rm IR} \approx 3\times10^{11}~L_{\odot}$, implying a star formation rate of $\sim 50~M_{\odot}~{\rm yr}^{-1}$) is highlighted for which a 50~$\mu$G magnetic field strength ({\it dashed}-line) is also shown.    
At the suggested depth for the SKA, it will detect galaxies forming stars at rates $\gtrsim 25~M_{\odot}~{\rm yr}^{-1}$) at all $z$ if present; 
this includes galaxies lying in the shaded region, indicating the expected flux density range for galaxies in the $z\sim$ 4, 5, 6, 7, and 9 UV luminosity function studies of \cite{rb07,rb08}.  
{\it Right~(b)-} The expected rest-frame IR/radio ratios for a galaxy having $q_{\rm IR}=2.64$ at $z=0$ as IC losses off of the CMB become increasingly important as a function of redshift along with data from a number of studies in the literature.  
Each track corresponds to a different internal magnetic field strength for the galaxy.  
As $U_{\rm CMB} \gg U_{B}$, the IR/radio ratio approaches the limit where only thermal (free-free) radiation contributes to the observed radio continuum emission.  
The average local $q_{\rm IR}$ values (2.64~dex; {\it dot-dashed} line) and the $\pm 1~\sigma$ scatter ({\it dotted}-line) are shown.  
Filled in symbols indicate that an object is an SMG, i.e. the entire \cite{ak06}, \cite{ev07},\cite{pc08},  \cite{ed09a,ed09b}, \cite{kc09}, \cite{vs09}, \cite{ak09}, and \cite{kk09}  samples, as well as half of the  \cite{ejm09a} sample.  
}
\label{fig-1}
\end{figure*}

Detecting moderately bright star-forming galaxies at redshifts beyond $z\sim3$ requires extremely sensitive continuum imaging only expected with the advent of the SKA. 
Figure \ref{fig-1}a shows the achieved point-source sensitivities for a number of existing 1.4~GHz radio continuum surveys including: 
VLA-COSMOS \cite{eva07}--5$\sigma_{\rm RMS}\approx 50~\mu$Jy, 
GOODS-N \cite{gm09}--5$\sigma_{\rm RMS}\approx 20~\mu$Jy, and 
the Deep SWIRE Field \cite{om08}--5$\sigma_{\rm RMS}\approx $13.5~$\mu$Jy.  
The expected point-source sensitivity of the EVLA after 300~hr ($\sim$1 Ms) is also plotted (5$\sigma_{\rm RMS}\approx 1.4~\mu$Jy).  

As already shown, a 5$\sigma$ point-source sensitivity of $\approx$40~nJy is required for the detection of a moderately bright LIRG 
forming stars at $\sim$50~$M_{\odot}~{\rm yr}^{-1}$, at $z\sim10$.  
However, it is unclear as to whether there will be a significant population of dusty star-forming galaxies at  these redshifts.  
At present, only rest-frame UV studies relying on the "dropout" technique are able to probe normal galaxies at such high redshifts.  
In Figure \ref{fig-1}a the estimated observed-frame 1.4~GHz flux densities, based on the results from the UV ($\approx$1600~\AA) luminosity function work of \cite{rb07,rb08}, are shown.  

These considerations suggest that the criterion of a point-source sensitivity of $\approx$40~nJy  
will only skim the brightest sources among this population of galaxies beyond $z\gtrsim7$.  
While this UV luminosity function work suffers dramatically from the effects of cosmic variance, 
it appears quite advantageous for the SKA sensitivity requirement be a factor of $\sim$2 better, namely a 5$\sigma$ point-source sensitivity of $\approx 20$~nJy (i.e. $\sigma_{\rm RMS} \approx 4$~nJy).

The sensitivity for the SKA is approximated using the radiometer equation such that \(\sigma_{\rm RMS} \approx 2k_{B}/\sqrt{2~BW~t_{\rm int}}~(A_{\rm eff}/T_{\rm sys})^{-1}\) where $k_{B}$ is the Boltzmann constant, BW is the bandwidth, $t_{\rm int}$ is the integration time, $A_{\rm eff}$ is the effective collecting area of the array, and $T_{\rm sys}$ is the system temperature.    
Assuming a bandwidth of  $\sim$1~GHz (comparable to what existing telescopes are delivering) and a reasonably long integration (i.e. 300~hr), I find a requirement for the SKA of  $A_{\rm eff}/T_{\rm sys} \sim 15000~{\rm m^{2}~K^{-1}}$ to reach this sensitivity (i.r. $\sigma_{\rm RMS} \approx 4~$nJy).
With these specifications, the SKA will be sensitive to all galaxies forming stars at a rate of $\gtrsim25~M_{\odot}~{\rm yr}^{-1}$.  
The $5\sigma_{\rm RMS} \approx 20$~nJy is also plotted in Figure \ref{fig-1}a.  

Given the above assumptions the SKA should be sensitive to a significant number of the high-$z$ galaxies reported in the UV luminosity function studies of \cite{rb07,rb08}.   
However, there are still large uncertainties in the current understanding of the sub-$\mu$Jy population of radio sources, and achieving such a high imaging dynamic range represents a significant technical challenge.  
The current generation of SKA pathfinders (e.g. \hbox{EVLA}, \hbox{ASKAP}, \hbox{MeerKAT}, and \hbox{APERTIF}) should usefully reduce the uncertainties in both the sub-$\mu$Jy radio source population(s) and high dynamic range imaging.

Finally, as the observed radio continuum emission from normal star-forming galaxies at high redshift  will likely be dominated by their thermal emission, this analysis does not require that dust be present in these galaxies for these sensitivity calculations to be valid.   
Specifically, this analysis requires their bolometric luminosities, and not necessarily their IR luminosities, be $\gtrsim10^{11}~L_{\odot}$ for detection.  
Thus, radio continuum emission from galaxies becomes the only way to find and quantify the star formation properties of galaxies unbiased by dust.  

\section{Summary}
In this proceedings article I have summarized the conclusions of \cite{ejmc09} which presented a predictive analysis for the expected evolution of the FIR-radio correlation versus redshift arising from variations in the CR electron cooling time-scales as IC scattering off of the CMB becomes increasingly important.   
In doing so, I have focussed on the value of deep radio continuum surveys in studies of star formation at high-$z$, particularly in the context of the SKA, finding the following:

\begin{enumerate}

\item Deep radio continuum observations at frequencies $\gtrsim$10~GHz using next generation facilities like the SKA will likely provide the most accurate measurements for the star formation rates of normal galaxies at high~$z$.   
The non-thermal emission from such galaxies should be completely suppressed due to the increased IC scattering off of the CMB leaving only the thermal (free-free) emission detectable; 
this situation may be complicated if `anomalous' microwave emission from spinning dust grains \cite{dl98a} in such systems is not negligible.  

\item For normal star-forming galaxies to remain on the local FIR-radio correlation at high redshifts requires extraordinarily large magnetic field strengths to counter IC losses from the CMB.  
For example, the magnetic field of a $z\sim6$ galaxy must be $\gtrsim$500~$\mu$G to obtain a nominal IR/radio ratio.  
Thus, galaxies which continue to lie on the FIR-radio correlation at high-$z$, such as the sample of radio-quiet QSO's, most likely have radio output which is dominated by an AGN, or the physics at work is simply unknown.  


\item To detect typical LIRGs ($L_{\rm IR} \gtrsim 10^{11}~L_{\odot}$) at all redshifts will require nJy sensitivities at GHz frequencies, specifically a 5$\sigma$ point-source sensitivity of $\approx$20~nJy (i.e. $\sigma_{\rm RMS} \approx 4$~nJy).  
Thus, for the SKA to achieve this sensitivity for a reasonably long (300~hr) integration necessitates that $A_{\rm eff}/T_{\rm sys} \sim 15000$~m$^{2}$~K$^{-1}$.  
At this sensitivity, the SKA will be sensitive to all galaxies forming stars at $\sim$25~$M_{\odot}~{\rm yr}^{-1}$.  
This includes a significant amount of sources included in the UV luminosity function work of \cite{rb07,rb08} between $4\lesssim z \lesssim 9$.  

\end{enumerate}

\end{document}